\documentclass[11pt]{article}
\usepackage{amsfonts,amsthm,amsmath,amssymb,slashed}
\usepackage[textwidth = 430 pt, textheight = 630 pt]{geometry}

\usepackage{amssymb}
\usepackage{amsmath}
\usepackage{amstext}
\usepackage{graphicx,epsfig}
\usepackage{epsfig}
\usepackage{verbatim} 
\usepackage{fancybox}
\usepackage{color}
\usepackage{ulem}
\usepackage{enumitem}
\usepackage{subfigure}
\usepackage{bbm}
\usepackage{parskip}

\newcommand{\Comment}[1]{{}}
\definecolor{MyDarkBlue}{rgb}{0.15,0.15,0.45}
\usepackage[linktocpage=true]{hyperref}
\hypersetup{
colorlinks=true,
citecolor=MyDarkBlue,
linkcolor=MyDarkBlue,
urlcolor=MyDarkBlue,
pdfauthor={Kurt Hinterbichler, Justin Khoury, Horatiu Nastase and Rogerio Rosenfeld},
pdftitle={Chameleonic inflation},
pdfsubject={hep-th}
}

\newcommand\ignore[1]{}
\def\one{{\,\hbox{1\kern-.8mm l}}}

\def\a{\alpha}\def\b{\beta}

\def\d{\partial}

\newcommand{\Cset}{{\,\,{{{^{_{\pmb{\mid}}}}\kern-.45em{\mathrm C}}}}}

\newcommand{\be}{\begin{equation}}
\newcommand{\bea}{\begin{eqnarray}}

\newcommand{\ee}{\end{equation}}
\newcommand{\eea}{\end{eqnarray}}

\providecommand{\lsim}{\lesssim}
\providecommand{\gsim}{\gtrsim}

\begin{document}

\renewcommand{\thefootnote}{\fnsymbol{footnote}}

\makeatletter
\@addtoreset{equation}{section}
\makeatother
\renewcommand{\theequation}{\thesection.\arabic{equation}}

\rightline{}
\rightline{}


\vspace{10pt}


\begin{center}
{\LARGE \bf{\sc Chameleonic inflation}}
\end{center} 
 \vspace{1truecm}
\thispagestyle{empty} \centerline{
{\large \bf {\sc Kurt Hinterbichler${}^{a,}$}}\footnote{E-mail address: \Comment{\href{mailto:khinterbichler@perimeterinstitute.ca}}{\tt khinterbichler@perimeterinstitute.ca}},
{\large \bf {\sc Justin Khoury${}^{b,}$}}\footnote{E-mail address: \Comment{\href{mailto:jkhoury@sas.upenn.edu}}{\tt jkhoury@sas.upenn.edu}},
{\large \bf {\sc Horatiu Nastase${}^{c,}$}}\footnote{E-mail address: \Comment{\href{mailto:nastase@ift.unesp.br}}{\tt nastase@ift.unesp.br}}
{\bf{\sc and}} 
{\large \bf {\sc Rogerio Rosenfeld${}^{c,}$}}\footnote{E-mail address: \Comment{\href{mailto:rosenfel@ift.unesp.br}}{\tt rosenfel@ift.unesp.br}}
                                                           }

\vspace{.5cm}

\centerline{{\it ${}^a$ Perimeter Institute for Theoretical Physics,}}
\centerline{{\it 31 Caroline St. N, Waterloo, Ontario, Canada, N2L 2Y5}}

\vspace{.3cm}
\centerline{{\it ${}^b$ 
Center for Particle Cosmology, Department of Physics and Astronomy,}}
 \centerline{{\it University of Pennsylvania, 209 South 33rd Street, }} \centerline{{\it Philadelphia, PA 19104, USA}}

\vspace{.3cm}
\centerline{{\it ${}^c$ 
Instituto de F\'{i}sica Te\'{o}rica, UNESP-Universidade Estadual Paulista}} \centerline{{\it 
R. Dr. Bento T. Ferraz 271, Bl. II, Sao Paulo 01140-070, SP, Brazil}}

\vspace{1truecm}

\thispagestyle{empty}

\centerline{\sc Abstract}

\vspace{.4truecm}

\begin{center}
\begin{minipage}[c]{380pt}
{\noindent We attempt to incorporate inflation into a string theory realization of the chameleon mechanism.  Previously, it was found that the volume modulus, stabilized by the supersymmetric potential used by Kachru, Kallosh, Linde and Trivedi (KKLT) and with the right choice of parameters, can generically work as a chameleon.  In this paper, we ask whether inflation can be realized in the same model.  
We find that we need a large extra dimensions set-up, as well as a semi-phenomenological deformation of the K\"{a}hler potential in the quantum region.
We also find that an additional KKLT term is required so that there are now two pieces to the potential, one which drives inflation in the early 
universe, and one which is responsible for chameleon screening at late times.  These two pieces of the potential are separated by a large flat desert 
in field space.  The scalar field must dynamically traverse this desert between the end of inflation and today, and we find that this can indeed occur under 
the right conditions.}
\end{minipage}
\end{center}

\vspace{.5cm}

\setcounter{page}{0}
\setcounter{tocdepth}{2}

\newpage

\renewcommand{\thefootnote}{\arabic{footnote}}
\setcounter{footnote}{0}

\linespread{1.1}
\parskip 4pt


\section{Introduction}
\ \ \ \ \ 
If our universe at high energies is governed by string theory, or any other higher-dimensional theory, then some of the extra dimensions have to be compactified and/or the standard model matter must be confined to a lower dimensional brane.  The size and shape of the compactification manifold, as well as the position of our brane world within the extra dimensions, appear to four dimensional observers as fundamental scalar fields.

Such fundamental scalars may play a primary role in the early universe as drivers of inflation.  Today, at late times, there is no sign of such a fundamental light scalar field, despite many experimental searches designed to detect the fifth forces that such a field would naively mediate~\cite{Fischbach:1999bc,Will:2005va}. If light scalars are present today, there must be some screening mechanism to explain why they have not been detected experimentally (see~\cite{Khoury:2010xi} for a review).  Experiments generally look for canonical linear scalars, and are performed near the Earth.  Screening mechanisms work by exploiting non-linearities which become important near the Earth, making the scalars difficult to detect.

Three broad classes of screening mechanisms have been developed from a phenomenological, bottom-up approach.  These are the chameleon mechanism~\cite{Khoury:2003aq,Khoury:2003rn,Gubser:2004uf,Upadhye:2006vi,Brax:2004qh,Brax:2004px,Mota:2006ed,Mota:2006fz,Brax:2007hi,Brax:2007vm,Brax:2009bk,Burrage:2007ew,Burrage:2008ii,Steffen:2009sc,Upadhye:2009iv}, which makes the effective mass of the scalar dependent on the local density of ordinary matter; the Vainshtein/galileon mechanism~\cite{Vainshtein:1972sx,ArkaniHamed:2002sp,Deffayet:2001uk}, which makes the scalar kinetic term large near matter sources; and the symmetron mechanism~\cite{Hinterbichler:2010es} (see also~\cite{Olive:2007aj,Pietroni:2005pv}), which exploits spontaneous symmetry breaking to weaken the coupling of the scalar near matter sources.  

A way to put these ideas on firmer theoretical footing is to realize them in string theory. Since galileons generally propagate superluminally on certain backgrounds, their UV completion cannot be a standard local quantum field theory or perturbative string theory~\cite{Adams:2006sv}. Chameleons and symmetrons, on the other hand, are described by canonical scalar fields with self-interaction potentials, and hence are compatible with a Wilsonian UV completion. Recently, some of us have presented a scenario for embedding the chameleon screening mechanism within supergravity/string theory compactifications~\cite{Hinterbichler:2010wu}. In this approach, the chameleon scalar field is identified with a certain function of the volume modulus of the extra dimensions.  The chameleon form for the potential and its coupling to matter arises from the Kachru, Kallosh, Linde and Trivedi (KKLT) construction \cite{Kachru:2003aw} for the superpotential for the volume modulus $\varrho$, albeit with a negative coefficient $\mathfrak{a}$ in the exponential $\sim e^{i\mathfrak{a}\varrho}$.  The main focus of~\cite{Hinterbichler:2010wu} was the late universe --- various constraints were found on the KKLT parameters which ensure that the scalar is screened from experiments today. These constraints, which are briefly reviewed in Sec.~\ref{modelreview}, lead to a large extra dimension scenario, with the most natural case consisting of 2 large extra dimensions (with the remaining 4 dimensions stabilized near the 10d fundamental scale).

In this paper, we extend this scenario to the early universe and ask whether is it possible that the volume modulus also drives slow-roll inflation in the early universe.  We find that this is indeed possible if we include the standard KKLT superpotential term $\sim e^{i\mathfrak{b}\varrho}$ with a positive $\mathfrak{b}$, together with a prefactor that guarantees inflation, but whose details do not influence much the subsequent evolution. We will consider a semi-phenomenological example for such a prefactor as a proof of principle. We find constraints on the parameters by demanding that inflation lasts sufficiently long and that the amplitude of density perturbations match observations. 

Thus in total, our model consists of a standard KKLT term, which is responsible for driving inflation in the early universe, and the $\sim e^{i\mathfrak{a}\varrho}$ term considered in \cite{Hinterbichler:2010wu}, which is responsible for chameleon screening at late times.  Only one of the terms is relevant at each epoch, so that the term responsible for inflation plays no role at late times, and the term responsible for chameleon screening plays no role at early times. The resulting model is economical, in that it involves a single scalar field that both drives inflation and acts as a chameleon today, and allows for the possibility that the inflaton field is directly observable experimentally today. (The idea that the inflaton still is a relevant field today appeared before, for instance in the quintessence form in \cite{Peebles:1998qn}, however we are not aware of this being considered before in the chameleon context.)  

Because the scalar field couples weakly to matter, reheating is inefficient and proceeds through gravitational particle production~\cite{Ford:1987de,Grishchuk:1990bj}. Inflation is therefore followed by a phase of kinetic domination. The scalar must exit the inflationary regime and reach the chameleon regime by the present time. It turns out that these two regions of the potential are separated by a large region with negligible potential  spanning many $M_{\rm Pl}^{-1}$ in distance --- a ``desert" --- which the scalar must dynamically traverse in order to accomplish this transition. In the ``minimal" scenario, in which kinetic domination is followed by a standard phase of radiation domination, we derive constraints
on the model parameters in order for the scalar to successfully traverse the desert and reach the chameleon region. We also consider an alternative scenario,
in which the universe temporarily becomes matter-dominated following kinetic domination, thanks to massive, unstable particle $X$ which decays well before big bang nucleosynthesis (BBN). The coupling of the volume modulus to the matter stress-energy drives the field towards the chameleon region of the potential. We find that a successful transition is achieved for a broad range of initial conditions and BBN proceeds as usual provided that the particle lifetime lies within the range $2.5 \times 10^{-3} \, \mathfrak{a}^{-4/3}~{\rm s} < \tau_X < 1~{\rm s}$. 

The paper is organized as follows. In Sec.~\ref{modelreview} we present the model, constraining both the chameleon and inflation regions. In Sec.~\ref{infcham} we analyze the 
consequences for inflation, deriving constraints on the relevant parameters, and discuss the reheating mechanism. In Sec.~\ref{rhdesert} we describe 
the physics needed for traversing the ``desert" between the inflation and chameleon region, and derive constraints on parameters required for a successful transition.
We summarize the results in Sec.~\ref{conclu}.

\section{The model}
\label{modelreview}

The model we consider is based on that of \cite{Hinterbichler:2010wu}, in which a scenario was proposed for embedding the chameleon within a supergravity/string theory compactification.  In this approach, the chameleon field is a function of the volume of the extra dimensions. The scenario uses the KKLT construction \cite{Kachru:2003aw} for the potential and the superpotential, with an important difference: the sign for the coefficient $\mathfrak{a}$ 
in the exponential for the superpotential, $Ae^{i\mathfrak{a}\varrho}$ is negative ($\mathfrak{a}<0$). While the original KKLT scenario has $\mathfrak{a}>0$, it was argued that even in the KKLT context one can  obtain extra terms with $\mathfrak{a}<0$, for instance by using gluino condensation on an extra D9-brane with magnetic flux as in \cite{Abe:2005rx}, or by imposing T-duality invariance on gaugino condensation superpotentials for tori compactifications as in \cite{Quevedo:1996sv}.
Another difference is that the scale relating the dimensionless $\varrho$ with the radius of extra dimensions was assumed in~\cite{Hinterbichler:2010wu} to be the 4d Planck scale $M_{\rm Pl}$, instead of the 10d Planck scale $M_{10}$, as in KKLT. Here we will instead follow the KKLT convention and work with 10d Planck scale.

Our goal in this work is to extend this framework to a more complete scenario, which allows the chameleon to also play the role of the inflaton in the early universe. 
In addition to the $Ae^{i\mathfrak{a}\varrho}$ term responsible for chameleon physics in the late universe, we also include in the superpotential the standard KKLT term $Be^{i\mathfrak{b}\varrho}$, with $\mathfrak{b}>0$,
to drive inflation at the fundamental Planck scale. To obtain a viable inflationary model, we will see that the coefficient $B$ must have a particular dependence on $\varrho$; while the assumed form
for $B(\varrho)$ is phenomenologically motivated and not governed by fundamental considerations, we will see that our model predictions are not sensitive to this choice. In this way our framework
unifies economically the idea of inflation and chameleon physics, both being defined by a single scalar: the volume modulus for the extra dimensions. 

In KKLT, the nonperturbative exponential superpotential arises for instance from Euclidean worldvolume instantons, such as Euclidean D3-branes wrapping
4-cycles of volume $V_4$ in the compact space. These contribute $e^{-S_{D3}}$ factors, where $S_{D3}$ is the Euclidean action of the wrapped branes~\cite{Witten:1996bn}. The physical 10d Newton's constant $\kappa_{10}$,\footnote{Note that usually one defines an unphysical $1/2\kappa_{10}^2$ and a physical $1/2\kappa^2=
1/(g_s^22\kappa_{10}^2)$, but we will call $\kappa_{10}$ the physical one, not to confuse with the 4 dimensional one.} 
the 10d Planck mass $M_{10}$, the string tension $\a'$, and the string coupling $g_s$ are related by 
\be
\frac{1}{2\kappa_{10}^2}\equiv \frac{M_{10}^8}{2}=\frac{1}{(2\pi)^7\alpha'^4 g_s^2}\,.
\ee
The D3-brane tension is
\be
\tau_3=\frac{\sqrt{\pi}}{\kappa_{10}}=M^4_{10}\sqrt{\pi}=\frac{2\pi M_s^4}{g_s} \,,
\ee
where $M_s\equiv (2\pi\sqrt{\alpha'})^{-1}$ is the 10d string scale. The dimensionless variable appearing in the exponent of the superpotential is $\varrho$,
whose imaginary part measures the volume $V_4$ of the 4-cycles wrapped by the Euclidean D3-branes:
\be
\sigma\equiv {\rm Im}\,\varrho=M_{10}^4V_4\sqrt{\pi} \,.
\label{sigmadef}
\ee

In terms of $\varrho$, our desired form of the superpotential is
\be
W(\varrho)=W_0+Ae^{-i|\mathfrak{a}|\varrho}+B(\varrho) e^{i\mathfrak{b}\varrho}\,,
\label{firsttry}
\ee
where $W_0$ is a constant. The first exponential, $\sim e^{-i|\mathfrak{a}|\varrho}$, has the opposite sign to the usual KKLT term, as mentioned earlier, and is responsible for the chameleon mechanism in the late universe.
The second exponential, with $\mathfrak{b}>0$, has the usual sign and will be responsible for driving inflation. Here $\mathfrak{b}$ can range anywhere from ${\cal O}(1)$ to $\ll 1$ (in KKLT, we can have 
$\mathfrak{b}\sim 1/N_c$ with $N_c$ a number of D7-branes). The chameleon part of the potential is reviewed in Sec.~\ref{chamreg}, whereas the inflationary part, in particular the $B(\varrho)$ prefactor,
is discussed in Sec.~\ref{infreg}.

\subsection{Chameleon Region}
\label{chamreg}

We first review the constraints from the chameleon part of the potential, neglecting the $B$ term in~(\ref{firsttry}):
\be
W_{\rm cham}(\varrho)\simeq W_0+Ae^{-i|\mathfrak{a}|\varrho} \,.
\label{chamsuperpot}
\ee
It is convenient to perform the analysis in terms of a new dimensionless
variable $R$, which measures the volume of the extra dimensions in 10 dimensional Planck units\footnote{An important difference with~\cite{Hinterbichler:2010wu} --- see Eq.~(3.6) there --- is that we have now defined $R$ in terms of $M_{10}$, consistent with the KKLT conventions, instead of $M_{\rm Pl}$.}:
\bea
\nonumber
{\rm d}s^2_D&=&R^2{\rm d}s_4^2+g_{\alpha\beta}{\rm d}x^\alpha {\rm d}x^\beta\,;\\
R&=& \frac{1}{\sqrt{V_6M_{10}^6}}\,.
\eea
The precise relation between $R$ and $\sigma$ depends on the hierarchy of scales in the extra dimensions. For concreteness, we will consider the case
where $n$ of the extra dimensions are large and of comparable size $r$, while the remaining $6-n$ dimensions are of the order of the fundamental scale $M_{10}^{-1}$.
In this case, 
\be
R = \frac{1}{\sqrt{V_6M_{10}^6}} = \frac{1}{(rM_{10})^{n/2}}\,.
\label{Rr}
\ee
Meanwhile, the 4d Planck scale upon dimensional reduction is given by\footnote{Here $M_{\rm Pl}$ denotes the reduced Planck scale: $M_{\rm Pl}\equiv (8\pi G_{\rm N})^{-1/2}\simeq 2.45\times 10^{18}$~GeV.}
\be
M_{\rm Pl}^2 \equiv M_{10}^8 V_6 = r^n M_{10}^{n+2}\,.
\label{Mpldimred}
\ee
For $\sigma$, we note that the leading instanton corrections are those with the largest action. These will be the instantons which wrap the largest cycles, {\it i.e.}, those wrapping all the large 
dimensions. If $n \geq 4$, then $V_4 = r^4$. If $n \leq 4$, on the other hand, then $V_4 = r^nM_{10}^{n-4}$. In other words, in this case the definition of $\sigma$ in~(\ref{sigmadef}) gives
\be
\sigma =\left\{\begin{array}{cl}
M_{10}^4r^4 \sqrt{\pi}  = \frac{2\pi M_s^4 r^4}{g_s}=R^{-\frac{8}{n}}\sqrt{\pi}
\hspace{20pt}&\text{for}\hspace{10pt} n \geq 4\,, \\ \\
M_{10}^n r^n \sqrt{\pi}   = \frac{2^{n/4}\pi^{(n+4)/8} M_s^n r^n }{g_s^{n/4}}=R^{-2}\sqrt{\pi}
\hspace{20pt}&\text{for} \hspace{10pt} n \leq 4 \,,
\end{array}\right.
\label{sigmaR}
\ee
where in the last step we have expressed $\sigma$ in terms of $R$. Up to an irrelevant factor of $\sqrt{\pi}$, which can be absorbed in a redefinition of parameters, we see that 
\be
\sigma=R^{-k}\,; \qquad \qquad  k = \left\{\begin{array}{cl}
\frac{8}{n} \hspace{20pt}&\text{for}\hspace{10pt} n \geq 4\,, \\ \\
2
\hspace{20pt}&\text{for} \hspace{10pt} n \leq 4 \,.
\end{array}\right.
\label{sigmaRbis}
\ee

As shown in our previous paper~\cite{Hinterbichler:2010wu}, and as we will review shortly, the resulting potential in the chameleon region
is dominated by a steep exponential part for values $R < R_\star$ and exhibits a minimum at $R_{\rm min}$. (This minimum is originally an AdS minimum, but following KKLT we lift it to a suitable dS minimum via a
supersymmetry breaking term.) The value of the field today is close to $R_*$, hence the constraint from laboratory experiments is naturally expressed in terms of $R_*$.
In~\cite{Hinterbichler:2010wu}, we obtained the condition 
\be
\mathfrak{a}R_*^{-k}\gsim 10^{30}\,.
\label{labconstr}
\ee
As shown in~\cite{Hinterbichler:2010wu}, this translates into a bound on the chameleon Compton wavelength at atmospheric density of $m_{\rm lab}^{-1} \lsim {\rm cm}$,
while in the solar system the bound is $m_{\rm solar}^{-1} \lsim 10^5$~km. 

Using~(\ref{Rr}), the above inequality translates into a bound on the present size of the large extra dimensions $r_*$ and the fundamental Planck scale $M_{10}$. Combining with the definition of $M_{\rm Pl}$ given in~(\ref{Mpldimred}), we can constrain $r_*$ and $M_{10}$ separately. The result is clearly sensitive to the number
$n$ of large extra dimensions. If $n=6$, for instance, corresponding to all extra dimensions being large, we find
\bea
\nonumber
M_{10}& \lsim & 25\mathfrak{a}^{3/4}~{\rm keV} \,; \\
r_* &\gsim & \frac{100}{\mathfrak{a}}~\mu{\rm m} \qquad \;\;\; (n = 6)\,.
\label{n6}
\eea
With $\mathfrak{a}\sim {\cal O}(1)$, $M_{10}$ is clearly ruled out by particle colliders, which preclude the existence of Kaluza-Klein modes with mass up to at least $\sim$TeV.

Repeating this exercise for different number of large extra dimensions, we find that the only allowed possibility is $n = 2$.
In this case,~(\ref{labconstr}) implies 
\bea
\nonumber
M_{10}&\lsim & 2.5\mathfrak{a}^{1/2}~{\rm TeV}\,; \\
r_* & \gsim &  \frac{100}{\mathfrak{a}}~\mu{\rm m} \qquad \;\;\; (n = 2)\,,
\label{n2}
\eea
which are just at the experimental limit with $\mathfrak{a}\sim {\cal O}(1)$.\footnote{When considering inflation later in the paper, we will find that $\mathfrak{a}\gg 1$ is
necessary for the field to successfully traverse the desert and reach the chameleon region. In this case, it may be
possible for $n > 2$ scenarios to be phenomenologically viable. For concreteness, however, we henceforth focus on $n=2$, since this is the
only case which allows $\mathfrak{a}\sim {\cal O}(1)$.} Although one might naively think that the constraints are even weaker for $n=1$, this is not the case --- the bound on $M_{10}$ remains the same, but $r_*$ must be absurdly large. (Unlike the Arkani-Hamed-Dimopoulous-Dvali (ADD) scenario~\cite{ArkaniHamed:1998nn,ArkaniHamed:1999gq}, 
where $n > 2$ is generically allowed, in our case the chameleon constraints only allow $n=2$, as~(\ref{n6}) illustrates.)

We are therefore led to consider a scenario with $n=2$ large extra dimensions, with the remaining 4 extra dimensions
of size at the fundamental scale $M_{10}$. To avoid conflicting with collider experiments, the standard model fields must be confined to a brane,
with the two large extra dimensions extending transverse to the brane. In other words, we need to live at least on a $D7$-brane, if not a $Dp$-brane with $p<7$.
The matter on the brane will couple to the induced metric on the brane, 
\be
g_{\mu\nu}^{\rm brane}=g_{MN}\d_\mu X^M\d_\nu X^N \,,
\ee
where $X^M$ are the embedding coordinates into the 10 dimensional space-time.
For simplicity, we demand that the brane is situated at a fixed point in these transverse extra dimensions, so there are no dynamical fluctuations for the brane position. (While not necessary, this choice avoids
dealing with the additional light scalars describing the brane position.) In this case, $g_{\mu\nu}^{\rm brane}$ is just the 10d metric $g_{MN}$ restricted to 
the brane position.  Hence $g_{\mu\nu}^{\rm brane}$ is still the same $g_{MN}$ Jordan frame metric. 

We are now in a position to specify the K\"{a}hler potential, which will define the canonical scalar field in terms of $\sigma$. 
In the perturbative region of the chameleon, we can use the tree-level K\"{a}hler potential found in string theory for the
volume modulus\footnote{Our form for $K$ can be justified as follows. If all the 6 dimensions are large, then 
\be
K=-3M_{\rm Pl}^2\ln[-i(\rho-\bar\rho)] \,.
\ee
When reducing on $n$ large dimensions down to d=4, we obtain in Einstein frame
\be
\frac{M_{10}^{2+n}}{2}\int {\rm d}^{4+n}x \sqrt{g^{(4+n)}}R^{(4+n)}=\frac{M_{\rm Pl}^2}{2}\int {\rm d}^4x \sqrt{g^{(4)}}\left[R^{(4)}-n\left(\frac{n}{2}+1\right)g^{\mu\nu}\d_\mu \psi\d_\nu \psi\right]\,,
\ee
where $r=e^\psi$, hence $\rho\propto ie^{n\psi}$ for $n\leq 4$. For $n=2$, this is obtained from the stated K\"{a}hler potential.}, 
\be
K(\varrho,\bar\varrho)\simeq -2M_{\rm Pl}^2\ln[-i(\varrho-\bar\varrho)]\,. \label{pertkahler}
\ee
The corresponding kinetic term for $\sigma$ is
\be
\frac{1}{2\sigma^2}M_{\rm Pl}^2(\d \sigma)^2\,,
\ee
hence the canonical scalar field is identified as
\be
\phi= M_{\rm Pl}\ln \frac{\sigma}{\sigma_*}\,,
\label{phicanonical}
\ee
where we have set $\phi = 0$ at the present time.

The supersymmetry potential that results from the superpotential~(\ref{chamsuperpot}) and the K\"{a}hler potential in (\ref{pertkahler})
is
\be
V_{\rm SUSY}(\sigma)=\frac{1}{2M_{\rm Pl}^2}\left[A^2\mathfrak{a}^2e^{2|\mathfrak{a}|\sigma}-\frac{2A|\mathfrak{a}|}{\sigma}e^{|\mathfrak{a}|\sigma}\Big(W_0+Ae^{|\mathfrak{a}|\sigma}\Big)-\frac{1}{2\sigma^2}\left(W_0+Ae^{|\mathfrak{a}|\sigma}\right)^2\right]  \,.
\label{poten}
\ee
We have the relation $\sigma = 1/R^2$, which follows from~(\ref{sigmaRbis}) for $n=2$. This potential has an AdS minimum at 
\be
\sigma_{\rm min} =\frac{1}{R_{\rm min}^2} \approx \frac{1}{|\mathfrak{a}|} \ln\frac{W_0}{A} \,,
\label{sigminW0}
\ee
which results from the combination of several exponentials in~(\ref{poten}). Meanwhile, $|\mathfrak{a}|(\sigma-\sigma_{\rm min}) \gsim 1$, the potential is well approximated by the leading exponential.
This defines the value $R_*$ (or $\sigma_*$) introduced earlier. Specifically, for $R \lsim R_*$ ($\sigma\gsim \sigma_*$), the potential is a steep exponential, $V(\sigma) \sim e^{2|\mathfrak{a}|\sigma}$.

\begin{figure}
   \centering
   \includegraphics[width=4.0in]{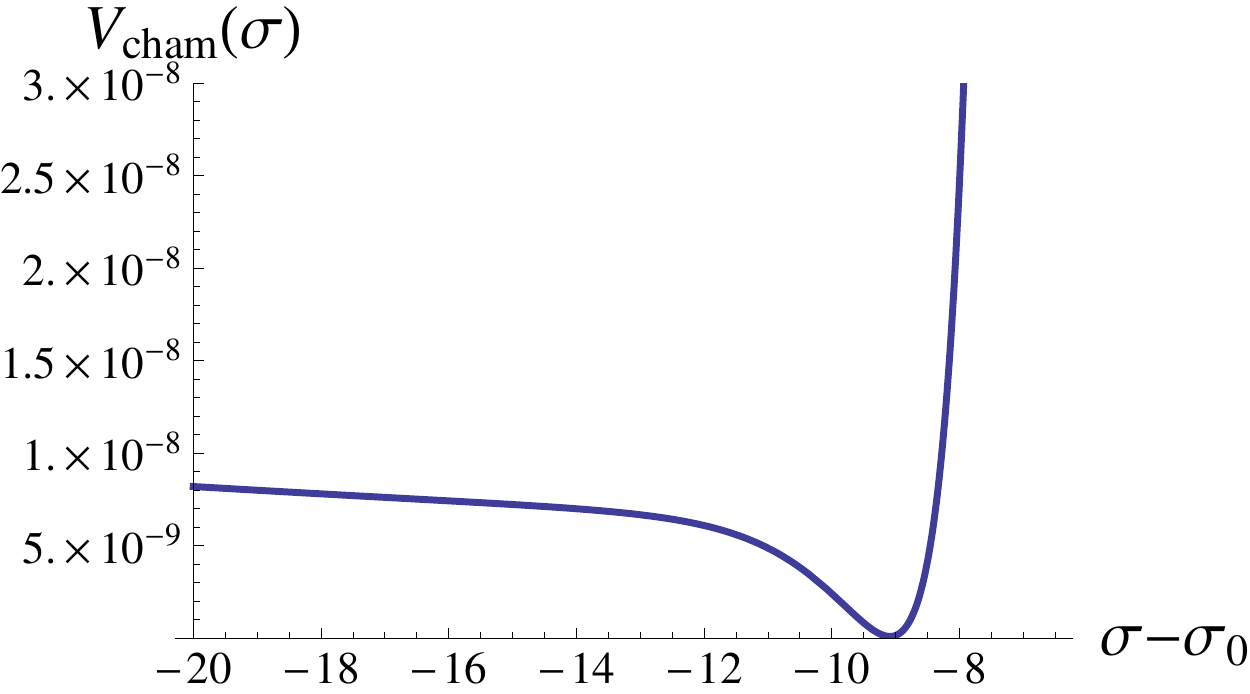}
   \caption{The potential (in Planck units) in the chameleon region consists of an exponentially steep part for large $\sigma$, with minimum at $\sigma = \sigma_{\rm min}$. See the main text for details.}
      \label{potentialsketch}
\end{figure}

As mentioned earlier, we have yet to supplement~(\ref{poten}) with a symmetry-breaking term to lift the AdS minimum
to a dS minimum with the right cosmological constant. Following KKLT, this is achieved by introducing an antibrane,
which contributes to the potential:
\be
V_{\rm SUSYbreak} =  \frac{D}{\sigma^2}\,.
\label{antibrane}
\ee
Since, for the parameters listed in~(\ref{potparams}), the depth of the AdS minimum
$V_{\rm today} \sim -10^{-150}M_{\rm Pl}^4$ is negligible compared to the observed positive cosmological constant, the antibrane contribution
is to a good approximation constrained to match the observed value $D/\sigma_*^2=10^{-122}M_{\rm Pl}^4$. With $\sigma_*\sim 10^{30}/\mathfrak{a}$, this fixes
$D\sim 10^{-62}\mathfrak{a}^{-2}M_{\rm Pl}^4$.

The total potential in the chameleon region, $V_{\rm cham}(\sigma) = V_{\rm SUSY}(\sigma)+ V_{\rm SUSYbreak}(\sigma)$, is sketch in Fig.~\ref{potentialsketch}.
For illustrative purposes, we chose the numerically-friendly values (in Planck units)
$|\mathfrak{a}| = 1$, $W_0 = 10^{-2}$, $\sigma_0 = 10^2$, $A = e^{-|\mathfrak{a}|\sigma_0}$ and $D = 0.78\,W_0^2$.

The chameleon potential is constrained by various phenomenological considerations, detailed in~\cite{Hinterbichler:2010wu}.  
For concreteness, we will adopt the limiting value of $R_* \simeq 10^{-15}\sqrt{\mathfrak{a}}$ allowed by~(\ref{labconstr}) --- recall that $k=2$ for the case $n=2$ of interest ---
corresponding to
\be
\sigma_* \simeq \frac{10^{30}}{\mathfrak{a}} \,. 
\label{sigma*maxvalue}
\ee
The parameters of the superpotential are then fixed as\footnote{Although some of the assumptions in~\cite{Hinterbichler:2010wu} were different, such as having $n=6$ large extra dimensions instead of 2, the expression for the potential in terms of $\sigma$ is qualitatively
unchanged, hence the results of the phenomenological analysis directly carry over to the present case.}
\be
W_0\sim 10^{-30}\mathfrak{a}^{-3/2}M_{\rm Pl}^3 \,;\qquad  \log\left(\frac{M_{\rm Pl}^3}{A}\right) \sim 10^{30}\,.
\label{potparams}
\ee
(While this naively implies a super-exponentially small value for $A$, one must keep in mind that the combination of parameters in the potential is $Ae^{|\mathfrak{a}|\sigma_*} $,
with $\sigma_*\sim 10^{30}/\mathfrak{a}$, hence $Ae^{|\mathfrak{a}|\sigma_*}$ can be comparable to $W_0\simeq 10^{-30}M_{\rm Pl}^3$ at the minimum). 

Finally, in the Einstein frame in which we will be working, the chameleon field $\phi$ couples conformally to matter on the brane as $F(\phi) T^\mu_{\;\mu}$,
where $T_{\mu\nu}$ is the matter stress tensor. For non-relativistic matter, this reduces to $F(\phi) \rho_{\rm matter}$. The coupling function,
\be
F(\phi) \equiv \frac{R}{R_*} = e^{-\phi/2M_{\rm Pl}}\,,
\label{Fdef}
\ee
relates the ($D$-dimensional) Jordan frame with the (4-dimensional) Einstein frame.
(Note that the coefficient in the exponent depends on $n$; in particular, in the $n=6$ case we had a different exponent.)

\subsection{Inflationary Region}
\label{infreg}

We have seen that at present time, corresponding to the chameleon region of the potential, 2 of the extra dimensions must be large.
In the early universe, on the other hand, we envision that inflation takes place at energy near the 10d fundamental scale
$M_{10}$, with all extra dimensions at the fundamental scale: $r\sim M_{10}^{-1}$, or $\sigma \sim 1$. In this region, we can neglect the first exponential in~(\ref{firsttry})
and approximate the superpotential as 
\be
W_{\rm inf}(\varrho)\simeq B(\varrho) e^{i\mathfrak{b}\varrho}\,.
\label{firsttry2}
\ee
Furthermore, to obtain a sufficiently flat potential to drive inflation, we need to modify the K\"{a}hler potential~(\ref{pertkahler}) as well. 
This is in any case to be expected, since the tree-level K\"{a}hler potential should receive $\a'$ corrections in the stringy region $r\sim M_{10}^{-1}$. The resulting modifications
generally occur within the logarithm. For instance, KKLMMT~\cite{Kachru:2003sx} considered the effect of D-brane positions $\Phi$, giving
(in our case of 2 large dimensions)
\be
K=-2M_{\rm Pl}^2\ln\left(-i(\varrho-\bar\varrho)-\kappa(\Phi,\bar\Phi)\right) \,.
\ee
In general, perturbative corrections give polynomial corrections inside the log, while nonperturbative corrections involve 
exponentials (see, {\it e.g.},~\cite{Quevedo:1996sv}).

We now specify suitable forms for $B(\varrho)$ and $\kappa$ to achieve successful inflation. While the particular $B(\varrho)$ we will use is admittedly contrived, 
we will argue in Sec.~\ref{rhdesert} that our results are largely insensitive to the particular forms for $B(\varrho)$ and $\kappa$,  as long as the resulting potential
displays an inflationary plateau, drops to zero as governed by the KKLT exponential $e^{-2\mathfrak{b}\sigma}$, and is followed
by a wide ``desert". The important physics for our considerations is captured by the KKLT exponential $e^{-2\mathfrak{b}\sigma}$, and in this sense
we describe physics associated with the KKLT model. Therefore we can view the form of $B(\varrho)$ and $\kappa$ specified here as a proof of 
principle. Our analysis applies more generally to any nonperturbative deformation that gives this kind of inflation. 

A simple supergravity model that allows slow-roll inflation is~\cite{Kachru:2003sx,Yamaguchi:2011kg}
$K(\varrho,\bar\varrho)=M_{\rm Pl}^2\bar\varrho \varrho$ and $W(\varrho) \sim \varrho$, corresponding to $V(\sigma) \sim 1+\sigma^4+...$. 
Following this approach, we consider nonperturbative corrections to the K\"{a}hler potential of the form\footnote{Note that, since ${\rm Im}\varrho\sim V_4/
\a'^2$, the correction inside the log is $\sim (\a'^2/V_4)e^{-V_4^2/\a'^2}$ with respect to the leading term.}
\be
K(\varrho,\bar \varrho)=-2M_{\rm Pl}^2\ln\left[-i(\varrho-\bar\varrho)+\lambda e^{-\a\varrho\bar\varrho}\right] \,.
\label{kahler}
\ee
As desired, this reduces to~(\ref{pertkahler}) at large $\varrho$ and tends to $K(\varrho,\bar\varrho)\simeq 
2\a M_{\rm Pl}^2\bar\varrho \varrho$ at small $\varrho$. Since we are instead interested specifically in new inflation, where the potential
slowly drops from an initial value and then faster down to a flat plateau (``the desert"), we will need two additional factors in the superpotential: 
$i)$ a factor of $e^{-\b_1 \varrho^4}\sim e^{-\b \sigma^4}$ (setting $\varrho=i\sigma$) to win over the $e^{2\alpha\sigma^2}$ factor in the potential at large $\sigma$;
and $ii)$ a non-perturbative factor of $e^{-\b_2 e^{c/i\varrho}}\sim e^{-\b_2 e^{-c/\sigma}}$, which triggers the end of inflation. Both factors are to a good approximation
equal to unity during inflation (at small $\sigma$), and hence will be irrelevant for inflationary observables.\footnote{The non-analytic factor of $e^{-\b_2 e^{-c/\sigma}}$ is necessary to prevent the occurrence of a maximum whose height is higher than the inflationary plateau, thus preventing the field from classically reaching the chameleon region of the potential.}
We therefore choose $B(\varrho)$ of the form:
\be
B(\varrho)=Be^{-\b_1 \varrho^4 -\b_2 e^{c/i\varrho}}\varrho \,,
\label{Bvar}
\ee
where $B$ is a constant. We should stress that it is not clear how this form would arise in string theory.  
The polynomial terms can conceivably arise as perturbative corrections to the instanton term $e^{i\mathfrak{b}\varrho}$, but the prefactor $e^{-\b_1 \varrho^4 -\b_2 e^{c/i\varrho}}$ is hard
to justify. That being said, for concreteness we will assume this form, in order to have a specific working inflationary model. As mentioned earlier, any
nonperturbative modification that achieves inflation will work for our purposes, and we consider this form of $B(\varrho)$ just as a proof of concept.

The complete superpotential is then
\be
W(\varrho)=W_0+Ae^{-i|\mathfrak{a}|\varrho}+Be^{i\mathfrak{b}\varrho}e^{-\b_1 \varrho^4 -\b_2 e^{c/i\varrho}}\varrho\,.
\label{superpotential}
\ee
Our model is thus fully specified by the K\"{a}hler potential~(\ref{kahler}) and the superpotential~(\ref{superpotential}).
In the next Section, we will focus on the inflationary dynamics and derive constraints on the various parameters to 
achieve successful inflation.

\section{Inflationary Dynamics and Constraints}
\label{infcham}

In this Section we describe in more detail the inflationary epoch, which is assumed to occur near the 10d fundamental scale
$M_{10}$, with the extra dimensions of size $r\sim M_{10}^{-1}$ ({\it i.e.}, $\sigma \sim 1$). In this regime, the
superpotential is approximately given by~(\ref{firsttry2}), with $B(\varrho)$ given in~(\ref{Bvar}):
\be
W_{\rm inf} (\varrho)\simeq Be^{i\mathfrak{b}\varrho}e^{-\b_1 \varrho^4 -\b_2 e^{c/i\varrho}}\varrho  \,.
\label{superpotentialinf}
\ee
Since the real component of $\varrho$ is assumed to be stabilized~\cite{Kachru:2003aw}, we will set $\varrho = i \sigma$ henceforth.
Through trial and error, we have scanned different values of the parameters to seek regions in parameter space where inflation is possible.
We found that inflation can be achieved in the region
\be
\alpha\sigma^2\ll \mathfrak{b}\sigma\ll 1\,,
\label{bsigll1}
\ee
provided that the model parameters satisfy
\be
\a\lambda^2\gg 1\,;\qquad \mathfrak{b}\lambda \gg 1\,.
\label{paramapprox}
\ee
The condition $\mathfrak{b}\sigma\ll 1$ is intuitively clear, since the superpotential depends exponentially on this combination. The other conditions will
be motivated below. We should stress that the above conditions are by no means necessary. Inflation may well be possible for other parameter values,
but among various possible simplifying approximations, this is the only one we have found to be compatible with inflation.

Starting with the kinetic term, the field-space metric that derives from the K\"{a}hler potential~(\ref{kahler}) is
\bea
\nonumber
g_{\varrho\bar\varrho} &=&2M_{\rm Pl}^2 \bigg( \frac{1+\lambda^2\a e^{-2\a\varrho\bar\varrho}-i(\varrho-\bar\varrho)\lambda\a e^{-\a\varrho\bar\varrho}(1-\a\varrho\bar\varrho)}
{[-i(\varrho-\bar\varrho)+\lambda e^{-\a\varrho\bar\varrho}]^2}\bigg) \\
&\simeq & 2M_{\rm Pl}^2\a \,,
\label{Kahlermetricfull}
\eea
where in the last step we have set $\varrho = i \sigma$ and used the conditions $\alpha\sigma^2\ll 1$ and $\lambda \alpha^2\gg 1$
from~(\ref{bsigll1}) and~(\ref{paramapprox}), respectively. This simple form for the K\"{a}hler metric implies a linear relation between $\sigma$
and the canonically-normalized scalar field
\be
\phi \simeq 2\sqrt{\a}M_{\rm Pl}\sigma \,.
\ee

Meanwhile,  ignoring contributions from the $e^{-\b_1 \sigma^4 -\b_2 e^{-c/\sigma}}$ prefactor, the scalar potential in Einstein frame is approximately given by
\be
V_{\rm inf}(\sigma) \simeq  \frac{B^2}{2\a \lambda^2 M_{\rm Pl}^2}\Bigg\{ 1-4\mathfrak{b}\sigma\bigg( 1 + {\cal O}\left(\frac{1}{\mathfrak{b}\lambda}\right)\bigg) +7\mathfrak{b}^2\sigma^2\bigg( 1 + {\cal O}\left(\frac{1}{\mathfrak{b}\lambda}\right)\bigg) +...\Bigg\},
\label{Vinfapprox1}
\ee
where the ellipses include terms of ${\cal O}(\alpha^2\sigma^4)$, which by~(\ref{bsigll1}) are negligible compared to the ${\cal O}(\mathfrak{b}^2\sigma^2)$ terms we have kept.
The ellipses also include terms of higher-order in $\sigma$, which only become important after inflation.

With these approximations, we can constrain various phenomenological quantities:

\begin{itemize}

\item {\bf Slow-roll parameters}: The standard $\epsilon$ and $\eta$ slow-roll parameters are 
\bea
\nonumber
\epsilon &=& \frac{M_{\rm Pl}^2}{2} \left(\frac{{\rm d}\sigma}{{\rm d}\phi} \frac{V_{,\sigma}}{V}\right)^2 \simeq \frac{2\mathfrak{b}^2}{\alpha} \,; \\
\eta &=& M_{\rm Pl}^2\left( \frac{{\rm d}^2\sigma}{{\rm d}\phi^2} \frac{V_{,\sigma}}{V} + \left(\frac{{\rm d}\sigma}{{\rm d}\phi}\right)^2 \frac{V_{,\sigma\sigma}}{V}\right) \simeq \frac{7\mathfrak{b}^2}{2\alpha}\,.
\label{epseta}
\eea
The spectral tilt of the scalar power spectrum is red-tilted~\cite{LiddleLyth}:
\be
n_s -1 = -6\epsilon + 2\eta \simeq - \frac{5\mathfrak{b}^2}{\alpha}\,.
\label{tilt}
\ee
Meanwhile, the tensor-to-scalar ratio is given by~\cite{LiddleLyth} 
\be r = 16\epsilon \simeq 32 \mathfrak{b}^2/\alpha.\ee

\item {\bf Primordial amplitude}: The observed primordial amplitude constrains~\cite{wmap9} 
\be
\frac{H_{\rm inf}^2}{8\pi^2\epsilon M_{\rm Pl}^2} \simeq 2.4\times 10^{-9}\,,
\ee
where $H_{\rm inf}^2= V_{\rm inf}/3M_{\rm Pl}^2$. Since $V_{\rm inf}\simeq B^2/2\a \lambda^2 M_{\rm Pl}^2$, and substituting~(\ref{epseta}), we obtain
\be
B\simeq10^{-3}  \lambda \mathfrak{b} M_{\rm Pl}^3\,.
\label{Bcons}
\ee

\item {\bf Number of e-folds}: As mentioned earlier, by assumption the end of inflation is triggered by the $e^{-\b_2 e^{-c/\sigma}}$ factor. This factor becomes relevant
when the field reaches $\sigma_{\rm end-inf} \sim c/\log(2\beta_2)$. Numerically, we have found that this estimate is off by a factor of 2, hence a more accurate estimate is
\be
\sigma_{\rm end-inf} \simeq \frac{c}{2\log(2\beta_2)}\,.
\ee
The total number of e-folds,
\be
N \simeq 2\sqrt{\alpha} \int_0^{\sigma_{\rm end-inf}} \frac{{\rm d}\sigma}{\sqrt{2\epsilon}} \simeq \frac{\alpha}{\sqrt{2}\mathfrak{b}} \sigma_{\rm end-inf} \,,
\ee
must be at $\simeq 60$ to solve the standard problems. This imposes $\sigma_{\rm end-inf} > 60\sqrt{2} \mathfrak{b}/\alpha$. Meanwhile, for consistency $\sigma_{\rm end-inf}$
must be less than $1/\mathfrak{b}$, which is the value when the approximation~(\ref{bsigll1}) breaks down. Thus we have the allowed range
\be
60 \sqrt{2} \frac{\mathfrak{b}}{\alpha} <  \sigma_{\rm end-inf} < \frac{1}{\mathfrak{b}}\,.
\label{sigendrange}
\ee

\end{itemize}

\begin{figure}
   \centering
   \includegraphics[width=4.0in]{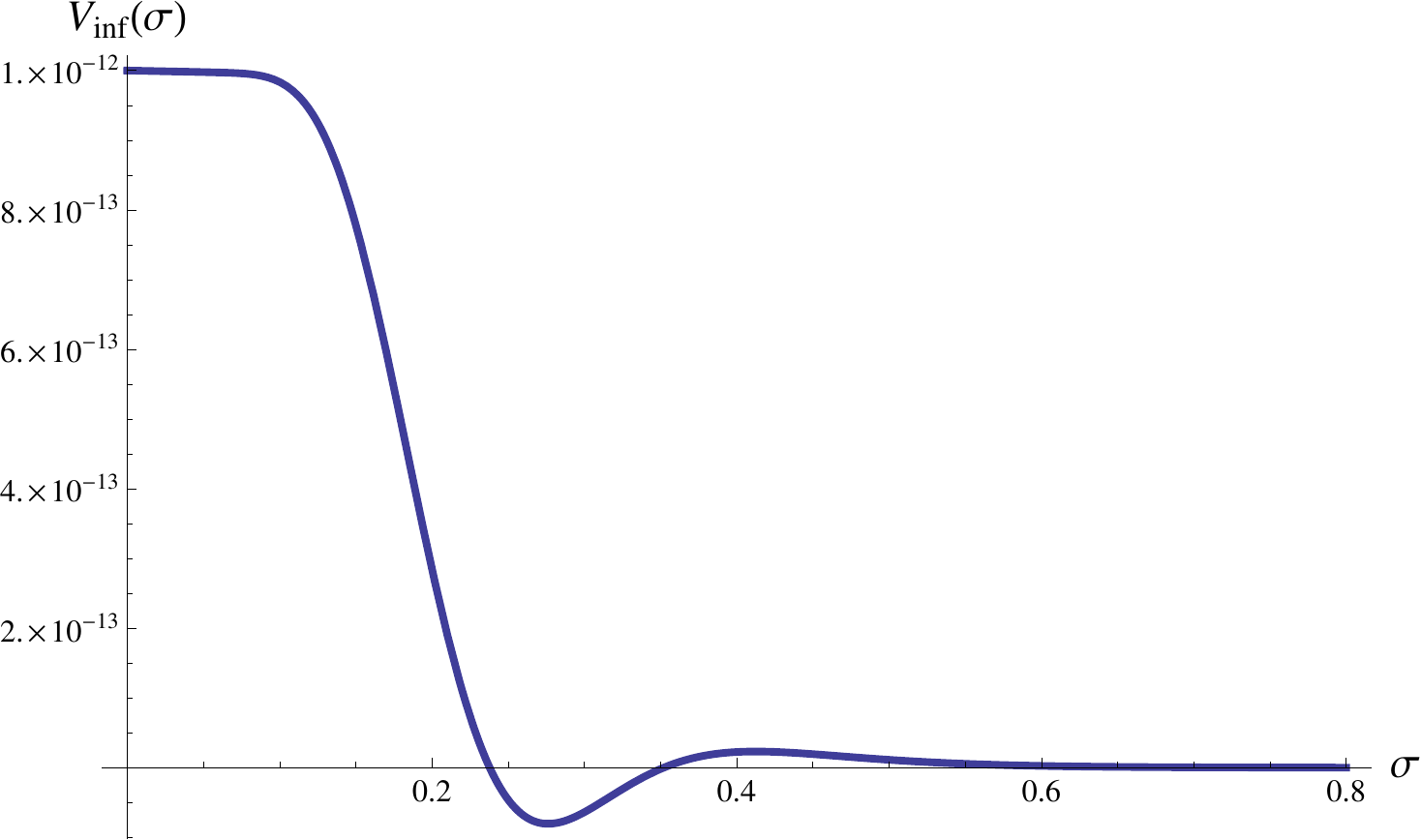}
   \caption{The region of the potential (in Planck units) relevant for inflation, for the parameter values~(\ref{fiducialinf}).}
         \label{infpotentialsketch}
\end{figure}

Figure~\ref{infpotentialsketch} shows the full supergravity potential that derives from the K\"{a}hler potential~(\ref{kahler})
and approximate superpotential~(\ref{superpotentialinf}), for the parameter values
\be
\alpha = \frac{1}{2}\,;~~ \mathfrak{b}=10^{-3}\,;~~c= 1\,;~~\beta_1 = 1\,;~~ \beta_2 = 10\,;~~\lambda = 10^2\,,
\label{fiducialinf}
\ee
with~(\ref{Bcons}) fixing $B \simeq 10^{-4}M_{\rm Pl}^3$. For these parameter values, the inflationary scale gives
\be
H_{\rm inf} \simeq 9\times 10^{-7}M_{\rm Pl} \simeq 2\times 10^{12}~{\rm GeV}\,.
\label{Hfid}
\ee
The inflationary plateau is followed by a shallow minimum and a flat region --- the ``desert". In the next Section, we will discuss how after inflation the scalar field manages to traverse this desert and reach the chameleon region by the present time. 

These parameter values satisfy~(\ref{sigendrange}), but
imply a tiny deviation from scale invariance, $n_s -1 \simeq 10^{-5}$, in tension with the measured value from the 9-year WMAP data~\cite{wmap9}: $n_s = 0.972 \pm 0.013$. We could not easily find alternative parameter values that would satisfy all constraints while yielding a desirable spectral tilt. The fiducial model therefore at best represents a proof of concept that an inflationary potential can be connected to a chameleon region within supergravity, but is not phenomenologically viable.

\subsection{Gravitational reheating}

As sketched in Fig.~\ref{infpotentialsketch}, inflation comes to an end when the field falls off the inflationary plateau. The potential energy
is converted into kinetic energy, and the field rapidly reaches the ``desert", where the potential is approximately zero over a
distance $\gg M_{\rm Pl}$ in field space. Since the scalar field couples weakly to matter, reheating is inefficient and proceeds through gravitational particle
production~\cite{Ford:1987de,Grishchuk:1990bj}. The reheating temperature is of order $H_{\rm inf}$, and the energy density in radiation is
\footnote{Note that, as in explained in \cite{Peebles:1998qn}, there is a factor $\alpha R^{3/4}$ in front of the expression for the reheating temperature,
with $\alpha$ a gauge coupling constant assumed in \cite{Peebles:1998qn} to be between $0.1$ and $0.01$ and $R=10^{-2}N_s$, with $N_s$ the number of scalars. 
In MSSM, $N_s=104$, and in string theory it can be larger. Also it is not clear what values the coupling $\alpha$ can take, it could in principle be close to 1.
Therefore we will assume that $\alpha R^{3/4}$ is of order 1 and drop it, though this prefactor is model dependent, and can take larger or smaller values 
as well.} 
\be
\rho_{\rm rad} \sim H_{\rm inf}^4 \,.
\label{rhoradinitial}
\ee
Thus most of the inflationary energy goes into scalar field kinetic energy
\be
\rho_{\rm kin} \simeq 3 H_{\rm inf}^2M_{\rm Pl}^2.
\label{rhokininitial}
\ee
By the end of reheating, the modification to the K\"{a}hler potential becomes negligible, and we are back to 
(\ref{pertkahler}).

As usual with gravitational reheating, we must ensure that the energy density in 4d massless gravitons is consistent with BBN bounds~\cite{Peebles:1998qn}.
At the end of inflation, the ratio of graviton and inflaton energy to the energy in matter fields is 
\be
f_{\rm end-inf} \equiv\frac{\rho_{\rm graviton} + \rho_{\rm inflaton} }{\rho_{\rm matter}}\bigg\vert_{a \simeq a_{\rm end-inf}} \sim \frac{3}{N_s}\,,
\ee
where $N_s$ is the number of scalar modes with masses below $H_i$. This translates at BBN to the ratio
\be
f_{\rm BBN} \equiv\frac{\rho_{\rm graviton} + \rho_{\rm inflaton} }{\rho_{\rm matter}}\bigg\vert_{a\simeq a_{\rm end-inf}} =\frac{3}{N_s}\left(\frac{{\cal N}_{\rm BBN}}{{\cal N}_{\rm th}}\right)^{1/3} \,,
\ee
where ${\cal N}_{\rm BBN} = 10.75$ is the effective number of spin degrees of freedom in equilibrium at temperature $T$,
while ${\cal N}_{\rm th}$ is the corresponding number at thermalization. Nucleosynthesis constraints impose $f_{\rm BBN} \lsim 0.07$,
which translates to ${\cal N}_{\rm th}\sim 10^2-10^3$. This is easily satisfied in our case: the MSSM has $N_s=104$ scalars,
while string theory has at least as many scalars. Thus this bound is satisfied in any case, though note that the bound is only valid in case there is no 
second reheating. In the second model for evolution towards BBN that we will present later, there is a second reheating, where the final amount of radiation 
present at BBN comes from the decay of a particle $X$. Then the original amount of gravitational radiation produced has been diluted away by the second 
reheating, and a negligible amount of gravitational radiation is produced now, since $X$ will couple more strongly to matter than gravitational strength.

A more constraining effect comes from the production of bulk gravitons from the brane, an effect first
considered in~\cite{ArkaniHamed:1998nn,ArkaniHamed:1999gq}. Once the extra dimensions have stabilized, 
there can in principle be evaporative cooling of the brane through bulk gravitons, which can compete with
the standard cooling from the FRW expansion. The condition in order for this not to happen is a bound
on the ``normalcy temperature", {\it i.e.}, the temperature at which we match onto the normal phase of
cosmological evolution at stabilization~\cite{ArkaniHamed:1998nn}:
\be
T_*\lsim 10^{\frac{6n-9}{n+1}}\left(\frac{M_{10}}{\rm TeV}\right)^{\frac{n+2}{n+1}}~{\rm MeV}   \,,
\ee
where $n$ denotes as before the number of large extra dimensions. Setting $n=2$ for the case of interest,
and substituting from~(\ref{n2}) $M_{10} \simeq 2.5\mathfrak{a}^{1/2}~{\rm TeV}$, where as before we have assumed the
limiting value~(\ref{sigma*maxvalue}), we obtain
\be
T_*\lsim 34\mathfrak{a}^{2/3}~{\rm MeV}\,.
\label{normalcy}
\ee

\subsection{Extra-dimensional volume during inflation}

We conclude this Section with a few brief remarks concerning the size of the extra-dimensional volume at the end of inflation.
In units of the 10d Planck scale $M_{10}$, the extra-dimensional volume reaches a maximal value of $V_6 \sim \sigma_{\rm end-inf}$
during inflation, which is always small. This problem also afflicts~\cite{Yamaguchi:2011kg}, and our modified superpotential
offers no improvement. Nevertheless, we would argue that this is simply an issue of initial conditions. There is certainly nothing wrong with 
the volume being of order unity in $M_{10}$ units --- it is usually assumed in standard implementations of inflation that 
quantum gravity effects stabilize the extra dimensions at the fundamental scale. The {\it stabilized value} of the volume should be greater
than or of order unity, but the volume at some arbitrary point in the $V(\phi)$ graph is not constrained. Effective field theory is still valid, since the {\it energy scale} of inflation
$V_{\rm in}$ is much less than unity in $M_{\rm Pl}$ units.

Note that the dilaton is assumed to be fixed in the KKLT-type construction, so there are no 
$g_s$ corrections, only $\a'$ corrections to the action. But $\a'$ corrections have already been incorporated in the action
for the scalar $\varrho$ in our semi-phenomenological approach --- see, {\it e.g.},~(\ref{kahler}) --- and these corrections
are indeed important in the inflationary region. One might also wonder about $\a'$ corrections to the gravitational action.
Such corrections to 4d gravity are small, $\a'R_4 \ll 1$, since the energy scale of inflation is sub-Planckian. Meanwhile,
corrections $\sim \a'R_6$ to 6d gravity on the compact space are implicitly taken into account in the action for $\varrho$. 

Nevertheless, let us pause and describe more explicitly what form such corrections might take. Corrections involving the 6d
Ricci scalar, $\sim \a'R_6^2$, must be small, for otherwise Einstein's equations would relate a large background $R_6$
to an unacceptably large background $R_4$.\footnote{This is why, for instance, the AdS$_4\times S^7$ background of 11d
supergravity cannot be modified to be phenomenologically acceptable.} For instance, a torus (with $R_6=0$) is acceptable,
but a sphere is excluded. For more general compact spaces, one also expects corrections involving the 6d Riemann tensor,
$\sim \a' R_{\mu\nu\rho\sigma}R^{\mu\nu\rho\sigma}$. However, in our framework we assume that only two compact dimensions
vary and can become large, while the remaining 4 are stabilized at the fundamental scale --- the standard geometries are  
$S^2$ (excluded) and $T^2$ (with $R_{\mu\nu\rho\sigma}=0$). Since the compact space is really 6 dimensional, small variations in the other 4 dimensions should induce small 
$\a'$ corrections for the action for $\varrho$, governing the volume of the two large dimensions, but these corrections have already been incorporated, as mentioned.\footnote{For $\varrho$ very close to zero, there will of course be other corrections from the appearance of light modes, such as string winding modes and brane wrapping modes on the shrinking cycles. But these can be neglected during the inflationary period.}

\section{Traversing the ``Desert"}
\label{rhdesert}

The length of the desert separating the inflationary plateau from the chameleon region is insensitive to the details of the potential.
It depends solely on the assumption of two large extra dimensions with fundamental scale at current epoch of $M_{10}$ and the fact that the 
K\"{a}hler potential for the volume modulus is given by the perturbative formula~(\ref{pertkahler}). 

Given the relation~(\ref{phicanonical}) between $\sigma$ and the canonical scalar field $\phi$, the distance to be traversed
is 
\be
\Delta\phi_{\rm desert} = M_{\rm Pl}\ln \frac{\sigma_\star}{\sigma_{\rm end-inf}} \simeq M_{\rm Pl}\ln \left(\frac{10^{30}}{\mathfrak{a}}\right)\,,
\label{desertlength}
\ee
where in the last step we have substituted the present value $\sigma_\star \simeq 10^{30}/\mathfrak{a}$ --- see~(\ref{sigma*maxvalue}) --- and estimated $\sigma_{\rm end-inf} \sim 1$. 
(For the parameter values~(\ref{fiducialinf}), for instance, we have $\sigma_{\rm end-inf} \simeq 0.2$.) As we will see below, such a large region is impossible
to traverse in the standard scenario of a chameleon coupled to radiation, unless one is willing to consider unnaturally large values of $\mathfrak{a}$. The initial kinetic energy~(\ref{rhokininitial}) rapidly redshifts away, and once the radiation component dominates, the scalar field is rapidly brought to a halt by Hubble friction. In order for the scalar field to successfully reach the chameleon region, we will assume
that the universe undergoes a temporary epoch of matter domination well before BBN, due to an unstable massive particle. The coupling of the scalar to this massive particle
will give the field the extra push required to reach its destination.  

Before proceeding, a few comments are in order. The required trans-Planckian separation is at first sight problematic in effective field theory,
since perturbative higher-dimensional operators like $\sim (\phi/M_{\rm Pl})^n \bar{\psi}\psi$ can become relevant for $\Delta\phi > M_{\rm Pl}$.
However, following KKLT we treat the supersymmetry breaking term due to the addition of an anti-brane in the chameleon region
as a small perturbation (at $\sigma\sim \sigma_*$), hence to a good approximation our setting is supersymmetric. In this context, 
the ``desert" acts as a supersymmetric valley for the minimum of the potential. General arguments suggest that such
corrections will not affect the supersymmetric locus of zero potential, {\it i.e.}, a desert remains a desert. 
(See, for instance,~\cite{BlancoPillado:2005fn} in the KKLT context, where it is noted that a supersymmetric extremum
valley of zero potential remains stable under quantum corrections.) 

More generally, as mentioned earlier this issue is a generic problem in scenarios with large extra dimensions. All we have used is the existence of large extra dimensions and 
the tree-level string theory K\"{a}hler potential for the volume modulus. It seems reasonable to assume that the potential for the volume modulus
in large extra dimensions remains small and monotonic until the stabilization point. So while the issue of potential corrections due to $\Delta\phi/M_{\rm Pl}>1$
is an important and subtle one (see {\it e.g.}, \cite{Kaloper:2011jz}), we will not address it further here, as it is outside the scope of this paper. 

\subsection{Kinetic-dominated phase}
\label{kindom}

Let us derive the distance traversed by the scalar during the kinetic-dominated phase that immediately follows inflation.
Since $\dot{\phi}\sim 1/a^3$, with its initial value fixed by~(\ref{rhokininitial}), we have
\be
\dot{\phi} = \sqrt{6}H_{\rm inf}M_{\rm Pl}\left(\frac{a_{\rm end-inf}}{a}\right)^3\,.
\label{dotphikin}
\ee
Meanwhile, during kinetic domination, $H = H_{\rm inf} (a_{\rm end-inf}/a)^3$. Combining these expressions, we obtain
\be
\frac{{\rm d}\phi}{{\rm d}\ln a} = \sqrt{6}M_{\rm Pl}\,,
\label{kindiff}
\ee
which implies a total distance traversed in the kinetic-dominated phase of
\be
\Delta\phi_{\rm kin} = \sqrt{6}M_{\rm Pl} \ln \frac{a_{\rm end-kin}}{a_{\rm end-inf}}\,,
\label{delphikin0}
\ee
where $a_{\rm end-kin}$ denotes the scale factor at the end of the kinetic dominated phase ({\it i.e.}, at kinetic-radiation equality). The latter can be determined straightforwardly
from the initial energy density in the kinetic and radiation components, given respectively by~(\ref{rhoradinitial}) and~(\ref{rhokininitial}):
\be
\frac{a_{\rm end-kin}}{a_{\rm end-inf}} \sim \frac{M_{\rm Pl}}{H_{\rm inf}} \,.
\label{aeqaendinf}
\ee
We therefore obtain
\be
\Delta\phi_{\rm kin} \simeq \sqrt{6}M_{\rm Pl} \ln  \frac{M_{\rm Pl}}{H_{\rm inf}}\,.
\label{Delphikin}
\ee
Note from~(\ref{aeqaendinf}) that the radiation density at the end of the kinetic-dominated phase is $\rho_{\rm end-kin} \sim H_{\rm inf}^8/M_{\rm Pl}^4$, with corresponding temperature of
\be
T_{\rm end-kin} \sim \frac{H_{\rm inf}^2}{M_{\rm Pl}}\,.
\label{Teq}
\ee
For the fiducial value~(\ref{Hfid}) $H_{\rm inf} \simeq 2\times 10^{12}$~GeV, this gives $T_{\rm end-kin} \simeq 1000~{\rm TeV}$.

Demanding that the field traverses the entire length of the desert during kinetic domination --- we will see shortly that the field moves by a relatively small amount during the subsequent radiation-dominated phase --- $\Delta\phi_{\rm kin}\gsim \Delta\phi_{\rm desert}$, we can combine~(\ref{desertlength}) and~(\ref{Delphikin})
to derive the following lower bound on $\mathfrak{a}$:
\be
\mathfrak{a}\gsim 10^{30} \left(\frac{H_{\rm inf}}{M_{\rm Pl}}\right)^{\sqrt{6}}\,.
\label{abound}
\ee
Since~(\ref{Teq}) in this case represents the temperature with which we reach stabilization,
and hence the temperature at the onset of the standard radiation-dominated phase, BBN requires
$T_{\rm end-kin} \gsim$~MeV, {\it i.e.},
\be
H_{\rm inf} \gsim 5\times 10^{7}~{\rm GeV} \,.
\label{HBBN}
\ee
On the other hand,~(\ref{Teq}) is bounded from above by the normalcy bound~(\ref{normalcy}): $T_{\rm end-kin} \lsim 34\mathfrak{a}^{2/3}~{\rm MeV}$.
This yields another lower bound on $\mathfrak{a}$, namely $\mathfrak{a}\gsim 0.6\cdot 10^{30} (H_{\rm inf}/M_{\rm Pl})^{3}$, but it is easy to check that 
this bound is weaker than~(\ref{abound}) for all values of $H_{\rm inf}$ satisfying~(\ref{HBBN}). In other words, once~(\ref{abound}) and~(\ref{HBBN})
are satisfied, the normalcy bound~(\ref{normalcy}) automatically follows.

For instance, for the fiducial value $H_{\rm inf} \simeq 2\times 10^{12}$~GeV, the bound~(\ref{abound}) gives
\be
\mathfrak{a} \gsim   10^{15} \qquad ({\rm with}\;\;H_{\rm inf} \simeq 2\times 10^{12}~{\rm GeV}\,;\;~T_{\rm end-kin} = 1000~{\rm TeV}) \,,
\label{aboundfid}
\ee
which is highly unnatural. For the lowest value allowed by BBN, $H_{\rm inf} \simeq 5\times 10^{7}~{\rm GeV}$, 
we instead obtain
\be
\mathfrak{a} \gsim 6\times 10^3 \qquad ({\rm with}\;\;H_{\rm inf} \simeq 5\times 10^{7}~{\rm GeV}\,;\;~T_{\rm end-kin} = {\rm MeV}) \,,
\label{abound1}
\ee
which is acceptable. To reiterate,~(\ref{abound})$-$(\ref{abound1}) apply in the ``minimal" model where the expansion history consists of inflation, followed by kinetic domination, followed by standard radiation-dominated cosmology. In Sec.~\ref{chamcoupled}, we will consider an alternative scenario where, after kinetic domination,
the universe is dominated by a massive relic $X$, which later decays into radiation. This alternative possibility will considerably lower the required value of $\mathfrak{a}$
in the fiducial case from $\mathfrak{a} \gsim  10^{15}$ to the more acceptable  $\mathfrak{a} \gsim 5\times 10^4$.

\subsection{Radiation-dominated phase}

Before moving on, however, we show that the field moves by a negligible amount during the subsequent radiation-dominated phase,
being quickly brought to a halt by Hubble friction. The kinetic energy continues to redshift according to $\dot{\phi} \sim 1/a^3$, hence
\be
\dot{\phi} = \sqrt{3}H_{\rm end-kin}M_{\rm Pl}\left(\frac{a_{\rm end-kin}}{a}\right)^3\,,
\ee
where the factor of $\sqrt{2}$ difference compared to~(\ref{dotphikin}) follows from the fact that the kinetic energy accounts for half of
the total energy density at equality. Meanwhile, during radiation domination the Hubble parameter evolves as $H  = H_{\rm end-kin} (a_{\rm end-kin}/a)^2$.
Instead of~(\ref{kindiff}), we obtain
\be
\frac{{\rm d}\phi}{{\rm d}\ln a} = \sqrt{3}M_{\rm Pl}\frac{a_{\rm end-kin}}{a}\,,
\label{raddiff}
\ee
with solution $\phi = - \sqrt{3}M_{\rm Pl} a_{\rm end-kin}/a + {\rm const}$. In the limit $a \gg a_{\rm end-kin}$, the distance traversed is
\be
\Delta\phi_{\rm rad} \simeq \sqrt{3}M_{\rm Pl}\,,
\label{Delphirad}
\ee
which does not help much in traversing the desert.

\subsection{Chameleon-coupled phase}
\label{chamcoupled}

To assist the scalar field in traversing the desert, we consider the possibility that there exists an unstable massive particle, $X$, which comes to dominate the energy density of the universe after kinetic domination. The scalar field couples to the energy density of $X$, and hence is driven to larger field values during $X$-domination.
While not necessary --- we have seen in Sec.~\ref{kindom} that it is possible for the field to traverse the entire desert during kinetic domination --- this alternative scenario
will broaden the phenomenologically allowed region of parameter space. In full generality, we study this problem numerically using the formalism of \cite{LopesFranca:2002ek},
assuming for concreteness that at the end of the kinetic-dominated era we have comparable amounts of kinetic, radiation and $X$ energy densities.

\begin{figure}[htb]
\begin{center}
\vspace{0.5cm}
\includegraphics[width=10.cm]{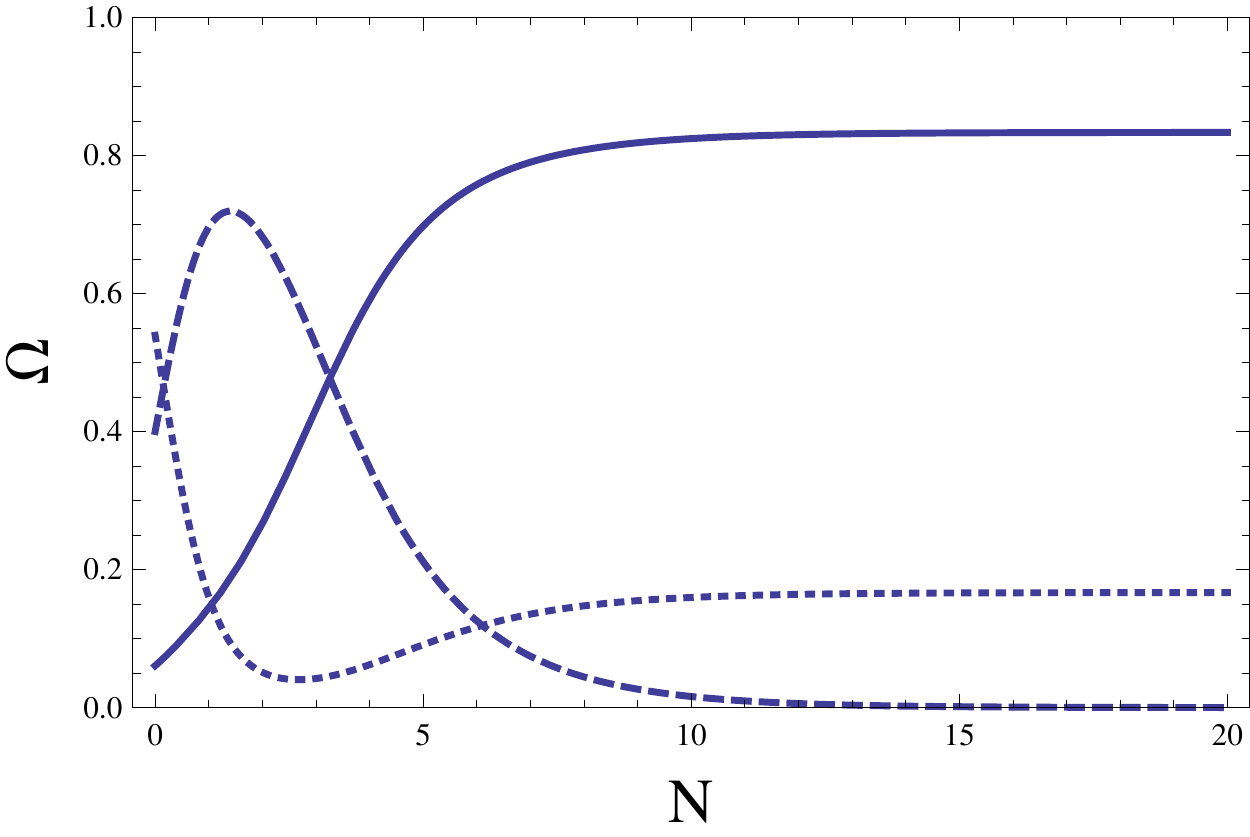}
\end{center}
\caption{Fractional densities $\Omega$ in units of the critical density for the chameleon-coupled matter $\Omega_X$ (solid line), chameleon field energy $\Omega_\phi$ (dotted line), 
and radiation $\Omega_{\rm rad}$ (dashed line), as a function of $N$ for $c = 1/2$.}
\label{omega}
\end{figure} 

Neglecting the potential energy, and using the perturbative relation~(\ref{phicanonical}) between $\sigma$ and the canonical scalar $\phi$,
the scalar equation of motion is
\be
\ddot{\phi} + 3 H \dot{\phi} = - \rho_X \frac{{\rm d}\ln F(\phi)}{{\rm d} \phi}\,,
\label{phicosmo1}
\ee
where $\rho_X \sim F(\phi)/a^3$. The coupling function is given by~(\ref{Fdef}), $F(\phi) = e^{-\phi/2M_{\rm Pl}}$, but
it will be instructive to consider the more general form
\be
F(\phi) =  e^{-c \phi/M_{\rm Pl}} \,,
\ee
where $c$ is a constant. (The case of interest is $c = 1/2$.) The Friedmann equation is
\be
3M_{\rm Pl}^2 H^2 = \rho_X + \frac{\dot{\phi}^2 }{2} + \rho_{\rm rad} \,.
\label{friedcosmo1}
\ee
It is convenient to change the time variable to the number of e-foldings, ${\rm d}N = {\rm d}\ln a$,
in terms of which~(\ref{phicosmo1}) becomes
\be
H^2 \phi'' + (3 H^2 + H H') \phi' =  - \rho_X \frac{{\rm d}\ln F(\phi)}{{\rm d} \phi}\,,
\ee
where $'\equiv {\rm d}/{\rm d}N$. The friction term $\sim (3 H^2 + H H') \phi'$ can be simplified by 
substituting the Friedmann equation~(\ref{friedcosmo1}) and the acceleration equation, 
\be
\frac{\ddot{a}}{a} = H' H + H^2 =  -\frac{1}{6 M_{\rm Pl}^2} \left( \rho_X + 2 \dot{\phi}^2 +  2 \rho_{\rm rad}  \right)\,,
\ee
to obtain
\be
H^2 \phi'' + \frac{1}{3 M_{\rm Pl}^2} \left( \frac{3}{2} \rho_X + \rho_{\rm rad}  \right) \phi' 
= - \rho_X \frac{{\rm d}\ln F(\phi)}{{\rm d} \phi}\,,
\label{phicosmo2}
\ee
where the Hubble parameter is determined by the Friedmann equation:
\be
3M_{\rm Pl}^2H^2 = \frac{\rho_X +  \rho_{\rm rad}}{1 - \frac{M_{\rm Pl}^2 \phi'^2}{6}}\,.
\label{friedcosmo2}
\ee

\begin{figure}[htb]
\begin{center}
\vspace{0.5cm}
\includegraphics[width=10.cm]{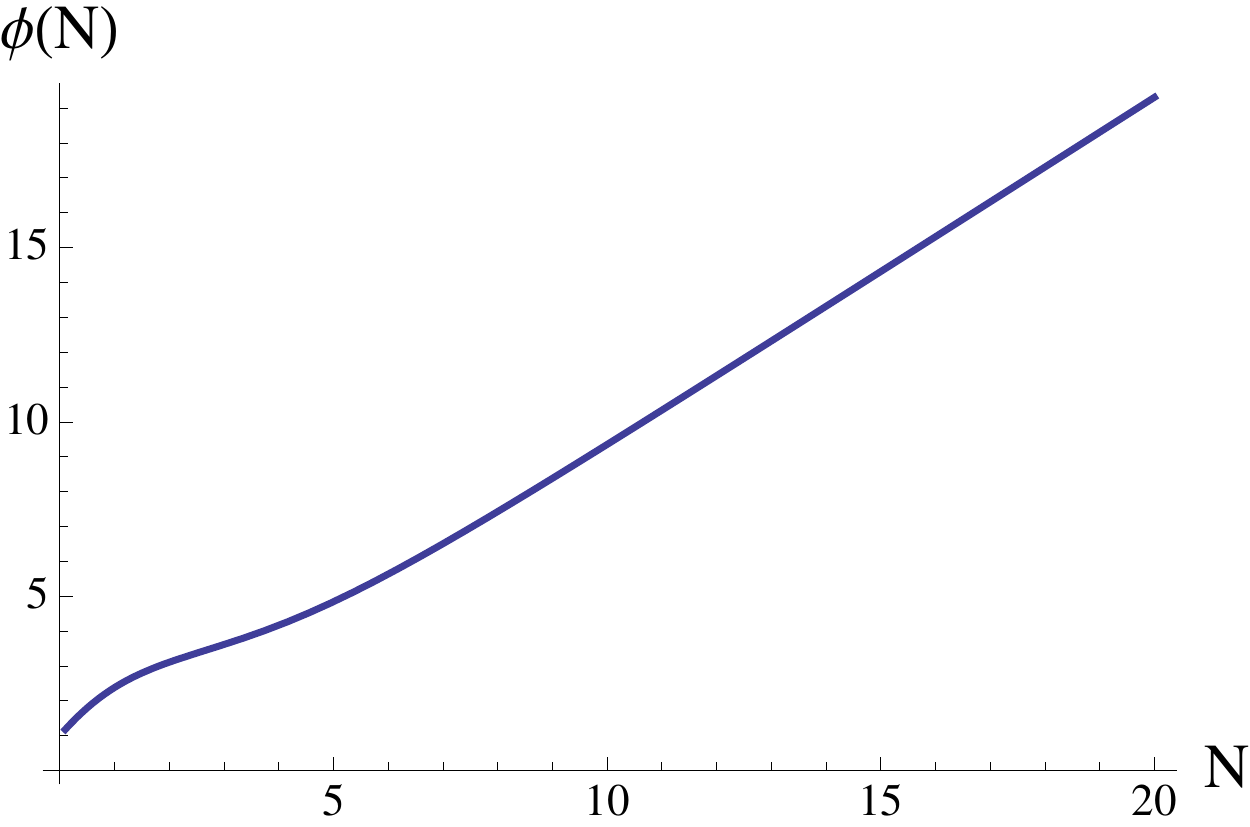}
\end{center}
\caption{Numerical solution for $\phi(N)$ (in Planck units) for an initial value of $\phi_{\rm i}= M_{\rm Pl}$, with $\Omega_X^{({\rm i})} = 0.1e^{-\phi_{\rm i}/2M_{\rm Pl}}$, $\Omega_{\rm rad}^{({\rm i})} =0.4$ and $\Omega_\phi^{({\rm i})} = 0.5$.}
\label{phiu}
\end{figure} 

Using the fact that $\rho_{\rm rad} \sim e^{-4N}$ and $\rho_X \sim F(\phi) e^{-3N}$,
and substituting~(\ref{friedcosmo2}) into~(\ref{phicosmo2}), we obtain an equation
for the scalar which only depends on $\phi$, its derivatives and $N$. This can be integrated
straightforwardly by specifying initial densities. Figure~\ref{omega} shows the resulting fractional energy densities,
$\Omega_\phi = \phi'^2/6M_{\rm Pl}^2$, $\Omega_X \equiv \rho_X/3H^2M_{\rm Pl}^2$ and
$\Omega_{\rm rad} \equiv \rho_{\rm rad}/3H^2M_{\rm Pl}^2$, for the fiducial case $c = 1/2$.

From the Figure, we see that the densities tend to a fixed-point behavior. (This was already anticipated in~\cite{TocchiniValentini:2001ty}
for the case of a constant potential and no baryons.) For $c <1/\sqrt{2}$, which includes the case of interest, the fixed point corresponds to
\be
\Omega_\phi = \frac{2}{3}c^2\;;\qquad \Omega_X = 1-\frac{2}{3}c^2\;;\qquad \Omega_{\rm rad} = 0\qquad\;\;\; {\rm for}~~c< \frac{1}{\sqrt{2}}  \,.
\label{fixedpt1}
\ee
For $c > 1/\sqrt{2}$, on the other hand, we found
\be
\Omega_\phi = \frac{1}{6c^2}\;;\qquad \Omega_X = \frac{1}{3 c^2}\;;\qquad \Omega_{\rm rad} = 1 - \frac{1}{2 c^2}\qquad\;\;\; {\rm for}~~c >  \frac{1}{\sqrt{2}} \,.
\ee
From~(\ref{phicosmo2}) and~(\ref{friedcosmo2}), the attractor solution for the field is
\be
\phi= \phi_{\rm i} + \frac{3c}{\frac{3}{2}+\frac{\Omega_{\rm rad}}{\Omega_X}}N =\left\{\begin{array}{cl}
\phi_{\rm i} + 2cM_{\rm Pl} N \hspace{20pt}&\text{for}\hspace{10pt} c<1/\sqrt{2} \,; \\ \\
\phi_{\rm i}  + \frac{M_{\rm Pl}N}{c} \hspace{20pt}& \text{for}\hspace{10pt} c>1/\sqrt{2} \,,
\end{array}\right.
\label{phiattract}
\ee 
where $\phi_{\rm i}$ is the initial field value. Note that for the case of interest, $c < 1/\sqrt{2}$, the attractor solution correspond to
$F(\phi) = e^{-c \phi/M_{\rm Pl}} \sim e^{-2c^2N}$, and
\be
\rho_\phi\,, \rho_X \sim e^{-(3+2c^2)N}\qquad \;\;\; {\rm for}~~c< \frac{1}{\sqrt{2}} \,.
\label{rhophirhoX}
\ee
(Since $c < 1/\sqrt{2}$, the radiation component $\rho_{\rm rad} \sim e^{-4N}$ redshifts faster than $\rho_\phi\,, \rho_X$ and is therefore driven
to zero, consistent with~(\ref{fixedpt1}).) For $c > 1/\sqrt{2}$, we instead find $\rho_\phi\,, \rho_X\,,\rho_{\rm rad} \sim e^{-4N}$,
that is, all components redshift like radiation, and the universe is effectively radiation-dominated. 

In our case, with $c = 1/2$, the attractor solution~(\ref{phiattract}) gives
\be
\Delta\phi =  M_{\rm Pl}  N \,.
\label{delphiattract}
\ee
Figure~\ref{phiu} shows the solution for $\phi(N)$ obtained numerically, which confirms the linear behavior~(\ref{delphiattract}) as the attractor solution.
Combining~(\ref{Delphikin}) and~(\ref{delphiattract}), and neglecting the contribution~(\ref{Delphirad}) from the radiation-dominated era,
the total distance traversed by the scalar field is
\be
\Delta\phi_{\rm tot} = \Delta\phi_{\rm kin} + \Delta\phi_X \simeq  \sqrt{6}M_{\rm Pl} \ln  \frac{M_{\rm Pl}}{H_{\rm inf}} + M_{\rm Pl} \ln \frac{a_{X-{\rm decay}}}{a_{\rm end-kin}}\,,
\label{delphitotal}
\ee
where $a_{X-{\rm decay}}$ is the scale factor when the $X$ particle decays. 
From~(\ref{desertlength}), we need $\Delta\phi_{\rm tot} \gsim M_{\rm Pl}\ln(10^{30}/\mathfrak{a})$ to successfully reach the chameleon region, which implies
\be
\frac{a_{\rm end-kin}}{a_{X-{\rm decay}}}\lsim 10^{-30} \mathfrak{a} \left(\frac{M_{\rm Pl}}{H_{\rm inf}}\right)^{\sqrt{6}}\,.
\label{aeqaXineq}
\ee

We next compute the reheating temperature once $X$ decays and its energy density is converted into radiation.
For simplicity, let us assume that $\rho_X$ is comparable to the kinetic energy and radiation at equality. That is, using~(\ref{Teq}),
$\rho_X(a_{\rm eq})  \sim H_{\rm inf}^8/M_{\rm Pl}^4$. Setting $c=1/2$ in~(\ref{rhophirhoX}), we obtain
\be
\rho_X(a_{X-{\rm decay}}) = \frac{H_{\rm inf}^8}{M_{\rm Pl}^4} \left(\frac{a_{\rm end-kin}}{a_{X-{\rm decay}}}\right)^{7/2}\,.
\ee
Assuming this energy all gets converted into radiation, the reheating temperature is
\be
T_{\rm final~reheat} =  \frac{H_{\rm inf}^2}{M_{\rm Pl}} \left(\frac{a_{\rm end-kin}}{a_{X-{\rm decay}}}\right)^{7/8}\lsim 10\,\mathfrak{a}^{7/8} \left(\frac{H_{\rm inf}}{M_{\rm Pl}}\right)^{2-\frac{7\sqrt{6}}{8}}~{\rm eV}\,,
\label{Trhbound}
\ee
where in the last step we have used~(\ref{aeqaXineq}). Numerically, $2- 7\sqrt{6}/8\simeq -0.14$, so the upper
bound is a slowly varying function of $H_{\rm inf}$. For our fiducial value of $H_{\rm inf} \simeq 2\times 10^{12}$~GeV, the inequality is 
$T_{\rm final~reheat} \lsim 70\,\mathfrak{a}^{7/8}$~eV. On the other hand, since BBN constrains $T_{\rm final~reheat} \gsim {\rm MeV}$, we obtain the following lower bound on $\mathfrak{a}$:
\be
\mathfrak{a} \gsim 5\times 10^4 \qquad ({\rm with}\;\;H_{\rm inf} \simeq 2\times 10^{12}~{\rm GeV})\,.
\label{aXfid}
\ee
Comparing with~(\ref{aboundfid}) derived in the minimal scenario, we see that including an intermediate $X$ phase greatly widens the
allowed range of parameters by lowering the value of the temperature at the onset of radiation domination.

The normalcy bound imposes the constraint $T_{\rm final~reheat} \lsim 34 \mathfrak{a}^{2/3}$~MeV, corresponding to
$H_{\rm final~reheat}^{-1} \gsim 2.5\times 10^{-3} \, \mathfrak{a}^{-4/3}$~s. This translates to a lower bound on the lifetime of $X$:
$\tau_X \gsim 2.5\times 10^{-3}\, \mathfrak{a}^{-4/3}$~s. Meanwhile, BBN requires $T_{\rm final~reheat} \gsim $~MeV, which
translates to $\tau_X \lsim 1$~s. The lifetime of $X$ must therefore lie within the window
\be
2.5\times 10^{-3} \, \mathfrak{a}^{-8/3}~{\rm s}\lsim \tau_X \lsim 1~{\rm s}\,.
\ee
For instance, in the fiducial case with the minimal value $\mathfrak{a} = 5\times 10^4$ allowed by~(\ref{aXfid}), this gives $3\times 10^{-9}
~{\rm s} \lsim \tau_X \lsim 1~{\rm s}$, which is a sizable window.

\subsection{Time evolution of $M_{10}$}

In Einstein frame, the 4d Planck scale $M_{\rm Pl}$ is by definition constant, but the 10d fundamental scale $M_{10}$ evolves in the time:
\be
M_{10}\sim \sigma^{-1/2} \sim e^{-\phi/2M_{\rm Pl}}\,.
\ee
For consistency, the typical energy scale of the dominant component must remain well below $M_{10}$ at all times. We check this epoch by epoch:

\begin{itemize}

\item During inflation, this is trivially satisfied: $M_{10}$ decreases from a value $\gg M_{\rm Pl}$ at early times to $\simeq M_{\rm Pl}$ at the end of inflation (since $\sigma_{\rm end-inf} \sim {\cal O}(1)$), while the inflationary scale is of course well below $M_{\rm Pl}$.

\item  During the kinetic domination,~(\ref{delphikin0}) implies that $M_{10} \sim a^{-\sqrt{3/2}}$, which therefore redshifts faster than
the temperature $T\sim a^{-1}$. Fortunately, kinetic domination ends before these scales have time to cross. Indeed, from~(\ref{Delphikin}) and~(\ref{Teq}), we infer
\be
M_{10}(a_{\rm end-kin}) \simeq \left(\frac{H_{\rm inf}}{M_{\rm Pl}}\right)^{\sqrt{\frac{3}{2}}} M_{\rm Pl} \gg T_{\rm end-kin}\,.
\ee

\item During $X$-domination,~(\ref{delphiattract}) implies $M_{10}\sim a^{-1/2}$, which redshifts more slowly than the radiation temperate $T\sim a^{-1}$.
It also redshifts more slowly than the typical energy scale $E_X\sim \rho_X^{1/4}$ of $X$ particles, which can be inferred from~(\ref{rhophirhoX}) with $c = 1/2$:
$E_X\sim a^{-7/8}$. Hence the $X$-dominated phase is completely safe.

\item Finally, during radiation domination, $M_{10}$ tends to a constant within a Hubble time as $\phi$ is brought to a halt, while the radiation temperature
of course continues to redshift.

\end{itemize}

\section{Conclusions}
\label{conclu}

In this paper we have extended the model of~\cite{Hinterbichler:2010wu}, which derived a chameleon scenario from a modified KKLT set-up, to 
include a complete cosmological evolution. In particular, our goal was to have the same scalar field, the volume modulus of the extra dimensions, to 
both drive the inflationary epoch at early times and play the role of a chameleon dark energy field at present time. Aside from economy, this
scenario opens up the exciting possibility of detecting the inflaton through present-day tests of gravity.

Borrowing the constraints from laboratory tests of gravity derived in~\cite{Hinterbichler:2010wu} , we were led to a large extra dimensions scenario,
with the most natural case having 2 large extra dimensions. To successfully obtain inflation, we modified in a semi-phenomenological way both 
the K\"{a}hler potential in the inflationary (stringy) region and the KKLT superpotential, $Be^{i\mathfrak{b}\varrho}$, with $\mathfrak{b}>0$
by letting $B\rightarrow B(\varrho)$. As a proof of principle, we focused on a particular --- and admittedly somewhat {\it ad hoc} --- form for $B(\varrho)$,
but we argued that the post-inflationary evolution is not sensitive to the details of the deformation. In the chameleon region of the potential, meanwhile,
we have the term $Ae^{i\mathfrak{a}\varrho}$, with $\mathfrak{a}<0$, used in \cite{Hinterbichler:2010wu}. We showed that the inflationary and chameleon regions
are separated by a large desert region, $\gg M_{\rm Pl}$ in field-space distance, where the potential energy is negligible.

We considered two scenarios in order for the scalar field to successfully traverse the desert after inflation.  The first is the ``minimal" scenario which does
not add any new ingredient.  After inflation, the universe enters a phase of kinetic domination, during which the scalar
field proceeds through the desert. Kinetic domination is generically long because reheating is inefficient and proceeds through
gravitational particle production. We derived constraints on the parameters, specifically $H_{\rm inf}$ and $\mathfrak{a}$, in order
for the field to reach the chameleon region by the onset of radiation domination. In this minimal scenario, therefore, the entire desert is
traversed during kinetic domination; the subsequent radiation-dominated phase is standard. 

In the second scenario, the universe after kinetic domination becomes dominated by a massive, unstable
relic $X$ which couples to the scalar field. This drives the scalar toward the chameleon region and helps it to traverse the desert. As a result, we found
that the allowed parameter space is broadened in this non-minimal scenario. Most of the $X$ matter must of course decay by BBN, which constrains its lifetime.
Otherwise, all experimental constraints on the cosmology can be satisfied with only the usual fine-tuning of large extra dimensions and of the cosmological constant,
as already noted in~\cite{Hinterbichler:2010wu}.

One issue we have not discussed in detail is how the field settles to the minimum of the potential in the chameleon region.
Let us focus on the case with $X$ matter, for concreteness. The attractor solution is due to the chameleon coupling, which creates a slope in the effective potential.
In its absence, the field would come to a stop due to Hubble friction before reaching the minimum. And since the attractor solution relies also on the fact that the matter energy density 
$\rho_X$ redshifts in time, thus lowering the slope of the effective potential, the field loses velocity gradually, with the kinetic energy $\rho_\phi$
being always of the order of the energy density of the chameleon-coupled matter $\rho_\chi$. But when reaching the minimum of the potential, the
energy density of the chameleon-coupled matter, and thus the kinetic energy as well, will be of the order of the potential energy --- the field
will gently settle at the minimum, being stopped by Hubble friction.

To improve on the scenario, it would be helpful to have a better-motivated inflationary potential. The explicit example studied here
serves as a proof of principle, as mentioned earlier, but is poorly motivated. We leave this to future work.

{\bf Acknowledgements} We thank Amanda Weltman for reading our manuscript and for discussions, and Albion Lawrence for discussions.
The work of J.K. is supported in part by NASA ATP grant NNX11AI95G, the Alfred P. Sloan Foundation and NSF CAREER Award PHY-1145525 (J.K.).
Research at Perimeter Institute is supported by the Government of Canada through Industry Canada and by the Province of Ontario through the Ministry of Economic Development and Innovation.  This work was made possible in part through the support of a grant from the John Templeton Foundation. The opinions expressed in this publication are those of the authors and do not necessarily reflect the views of the John Templeton Foundation (K.H.). The work of H.N. is supported in part by CNPQ grant 301219/2010-9.

\bibliography{cipaper18}
\bibliographystyle{utphys}

\end{document}